\documentclass[12pt]{article}
\usepackage{epsfig}
\usepackage{amssymb}
\newcommand{\be}{\begin{equation}}
\newcommand{\ee}{\end{equation}}
\newcommand{\bea}{\begin{eqnarray}}
\newcommand{\eea}{\end{eqnarray}}

\def\gev{\, \, {\rm GeV}}

\newcommand{\PYTHIA}{PYTHIA}

\begin{document}
\thispagestyle{empty}
\begin{flushright}
\texttt{RECAPP-HRI-2011-004}\\

\end{flushright}
\vskip 15pt

\begin{center}
{\Large {\bf Signals of an invisibly decaying Higgs in a scalar dark matter scenario: a study for the Large Hadron Collider}}
\renewcommand{\thefootnote}{\alph{footnote}}

\hspace*{\fill}

\hspace*{\fill}

{ \tt{ Kirtiman Ghosh$^{1}$, Biswarup Mukhopadhyaya$^{1}$ and
Utpal Sarkar$^{2,3}$
}}\\

\small {\em $^{1}$Regional Centre for Accelerator-based Particle Physics,\\
Harish-Chandra Research Institute, India.}\\
\small {\em $^{2}$Physical Research Laboratory, India.}\\
\small {\em $^{3}$McDonnell Center for the Space Sciences, \\
Washington University in St. Louis, USA.}\\

\vskip 40pt

%%%%%%%%%%%%%%%%%%%%%%%%%%%%%%%%%%%%%%%%%%%%%%%%%%%%%%%%%%%%%%%%%%%%%%%%%%%%%%%%%%%%%%%%%%%
{\bf ABSTRACT}\\
\vskip 0.5cm
\end{center}

We consider the collider phenomenology of a singlet Majoron model with softly broken lepton number. Lepton number is spontaneously broken when the real part of a new singlet scalar develops vacuum expectation value. With the additional soft terms violating lepton numbers, the imaginary part of this singlet scalar becomes a massive pseudo-Majoron which can account for the dark matter. In presence of the coupling of the pseudo-Majoron with the Standard Model (SM) Higgs, the SM Higgs mostly decays into a pair of pseudo-Majorons, giving rise to $E_T\!\!\!\!\!\!/~~$ signals at a hadron collider. Since the Higgs visible decay branching fractions get reduced in presence of this invisible decay mode, the bounds on the SM Higgs mass from the LEP and Tevatron experiments get diluted and the invisible decay channel of the Higgs become important for the discovery of low mass Higgs at the Large Hadron Collider.

\vskip 30pt

\section{Introduction}
The Standard Model (SM) of strong and electroweak interactions is a well-studied theory. The theoretical pillar of the SM is invariance with respect to the gauge group $SU(3)_C \times SU(2)_L \times U(1)_Y$ and spontaneous breakdown of the electroweak gauge group $SU(2)_L \times U(1)_Y$ to the $U(1)$ group corresponding to electromagnetism. We see phenomena which apparently are results of electroweak symmetry breaking (EWSB): the electroweak gauge bosons namely, the $W^\pm$ and the $Z$, get masses and so do the chiral fermions. However, the ultimate source of EWSB is still mysterious. In the framework of the SM, EWSB is achieved through a scalar $SU(2)$ doublet which couples to the gauge bosons via the covariant derivative, and to the fermions via Yukawa couplings. The vacuum expectation value (VEV) of this doublet becomes responsible for the masses of the $W^\pm$, $Z$-boson and the chiral fermions and one physical degrees of freedom, namely the {\em Higgs boson}, remains in the mass spectrum. The Higgs has so far successfully thwarted all attempts towards its detection. However, from all other measurements, namely the gauge couplings and the masses of gauge bosons and fermions, the couplings of the Higgs bosons to all SM particles are fixed so is the strength of its self-coupling once the mass and the VEV of the Higgs are known. Thus, the collider phenomenology of the SM Higgs boson is completely determined by just one undetermined parameter, namely the Higgs boson mass.

However, the correct theory of EWSB may be different from the SM, atleast in the sense that the Higgs sector can be richer. Many new models in this direction have been proposed during last two or three decades. The twin primary goals of the ongoing Large Hadron Collider (LHC) experiment are to understand the mechanism for electro-weak symmetry breaking (EWSB) as well as uncover any new dynamics beyond the SM that may be operative at the scale of a TeV or so.

The signature of the Higgs boson at the LHC has been extensively studied in the framework of the SM and its various extensions \cite{SMh1,SMh2}. However, there may still be possibilities leading to qualitatively different signatures for the Higgs boson. In particular, some extensions of the SM may contain a Higgs boson that can decay into stable neutral weakly interacting particles, therefore giving rise to invisible final states at collider experiments. Examples of this are Majoron models \cite{mj1,mj2,mj3,mj4,mj5} which are quite popular in the context of neutrino mass generation and leptogenesis. If neutrinos are Majorana particles, lepton number must be broken by the neutrino mass. In the simplest version of the Majoron model \cite{mj1} allowed by $Z$-decay data, global lepton number is spontaneously broken after a complex singlet scalar carrying lepton number develops a VEV. Right-handed neutrinos acquire Majorana masses through their Yukawa coupling with this singlet scalar. As a result of the spontaneous breaking of global lepton number, a massless neutral pseudoscalar, called the Majoron, remains in the spectrum. It can acquire mass and become a massive pseudo-Majoron in some variants of the singlet Majoron model \cite{smj1,smj2,us}. As for example, one can take into account the soft breaking of the global lepton number, which gives rise to a pseudo Goldstone boson. It was shown in Ref. \cite{us}, that such a massive pseudo-Majoron can account for the cold dark matter content of the universe. Moreover, the quartic coupling of this singlet scalar with the SM Higgs doublet opens up new decay modes for the Higgs boson, namely, the decay of Higgs boson into a pair of pseudo-Majorons. The pseudo-Majoron, being invisible and stable, gives rise to missing energy signature at colliders. In this article, we have studied the decay of Higgs boson in to a pair of pseudo-Majorons and its consequences in the context of the Large Hadron Collider (LHC).

At the LHC, the Higgs search strategies are modified considerably and in fact become more challenging in presence of invisible Higgs decay modes. This is in contrast to an $e^+e^-$ collider where, via the production channel $e^+e^- \to ZH$, even an invisible Higgs shows up as a resonance recoiling against a $Z$-peak \cite{ilch1,ilch2}.
%However, at $e^+e^-$ collider, the problem is not that much severe because here%, the dominant channel for the Higgs search is same for the SM Higgs and invisi%bly decaying Higgs: the Bjorken production channel, $e^+e^- \to Z^*H$, followed% by $Z^*\to \nu \bar \nu$, $H\to b \bar b$ for the SM decay and $Z^*\to q \bar %q$, $H\to$ invisible for the invisible decay \cite{ilch1,ilch2}. It is importan%t to note that unlike $e^+e^-$ colliders, missing energy is not measurable quan%tity at the LHC due to the lack of information about the longitudinal boost of %the center-of-mass frame. One has to consider the missing transverse momentum (%$p_T\!\!\!\!\!\!\!/$) instead.
In order to detect the signature of an invisibly decaying Higgs boson at the LHC, one still has to look for some associated production channel. The dominant production process there for a not-too-heavy Higgs  is gluon-gluon fusion ($gg \to H$). If the Higgs decays invisibly, one can look for the production of Higgs in association with a hard jet ($gg \to H+{\rm jet}$). The resulting signature would be then a high $p_T$ jet and missing transverse energy ($E_T\!\!\!\!\!\!/~$). However, this monojet in association with missing transverse energy signal is overwhelmed by the QCD background. The next dominant Higgs production channel is vector boson fusion ($qq \to qq V^* V^* \to qqH$). Here the final state will consist of two energetic high rapidity jets and missing-$E_T$ \cite{Eboli:2000ze,Datta:2001hv}. In this work, we have considered a relatively subdominant production channel, namely the production of the Higgs boson in association with a $W$ or a $Z$-boson ($q\bar q^{(\prime)}\to HZ(W^{\pm})$) followed by the leptonic decay of $W$ and $Z$-boson, where the leptons in the final state make the signal relatively cleaner \cite{dc,rg}.

This paper is organized as follows. In the next section, we briefly discuss the pseudo-Majoron model and invisible decay of the Higgs boson. Section 3 contains the main characteristics of the signal and backgrounds and cuts chosen to enhance the signal to background ratio. Our numerical results are presented at the end of that section. We summarise and conclude in section 4.

\section{The Model}

In addition to the SM fields, the model \cite{us} under
investigation includes two types of additional singlet fields: one
is a scalar ($\xi$) and the other set contains three right-handed
neutrinos ($N_R$). Since SM leptons carry the lepton number $L=1$,
same lepton number is assigned to the right-handed neutrinos. On
the other hand, $L=-2$ is assigned to the singlet scalar $\xi$.
The part of the Lagrangian involving neutrinos can be written as:
\begin{eqnarray} \label{yukawa}
\mathcal{L}_{Y}^{}\supset - y_{\nu}^{}\bar{\psi}_{L}^{}\phi
N_{R}^{}-\frac{1}{2}h\xi\bar{N}_{R}^{c}N_{R}^{}+\textrm{H.c.}\,,
\end{eqnarray}
where, $\psi_L^{}$ and $\phi$, are the SM lepton and Higgs
doublets  respectively. The scalar potential is given by,
\begin{eqnarray}
\label{potential}
V(\xi,\phi)&=&-\mu_{1}^{2}\xi^{\dagger}_{}\xi+\lambda_{1}^{}(\xi^{\dagger}_{}\xi)^{2}_{}-\mu_{2}^{2}\phi^{\dagger}_{}\phi
+\lambda_{2}^{}(\phi^{\dagger}_{}\phi)^{2}_{}\nonumber\\
&+&2\lambda_3^{}\xi^{\dagger}_{}\xi\phi^{\dagger}_{}\phi\,,
\end{eqnarray}
where $\lambda_{1,2}^{}>0$ and
$\lambda_3^{}>-\sqrt{\lambda_1^{}\lambda_2^{}}$ to guarantee the
potential bounded from below.

When the singlet scalar acquires a vacuum expectation value
(VEV), it breaks lepton number and gives Majorana masses to the
right-handed neutrinos. Spontaneous breaking of lepton number
would then give rise to an unwanted massless Majoron. For this
reason, soft breaking of the lepton number is taken into account
with the following term,
\begin{equation}
V_{soft}=-\mu_{3}^{2}(\xi^{2}_{}+\textrm{H.c.})/2.
\end{equation}
As a consequence, the massless Majoron becomes a massive
pseudo-Majoron. All other soft lepton number violating terms can
be forbidden by appropriate discrete symmetries such as a $Z_4$
symmetry, under which $\phi\rightarrow \phi\,,\quad\quad
\xi\rightarrow -\xi\,,\quad f\rightarrow i f\,$, where $f$ stands
for the SM fermions and the right-handed neutrinos.

We now expand the singlet scalar field in terms of its real and
imaginary components $\xi=(\sigma +i \chi)/\sqrt 2$, where the
real part $\sigma$ includes its VEV and the physical field,
while the imaginary part represents the pseudo-Majoron. We can
then rewrite the full scalar potential in the following form:
\begin{eqnarray}
\label{potential_f}
V&=&-\frac{1}{2}(\mu_1^2+\mu_3^2)\sigma^2_{}+\frac{1}{4}\lambda_1^{}\sigma^4_{}-\frac{1}{2}(\mu_1^2-\mu_3^2)\chi^2_{}\nonumber\\
&&+\frac{1}{4}\lambda_1^{}\chi^4_{}-\mu_{2}^{2}\phi^{\dagger}_{}\phi
+\lambda_{2}^{}(\phi^{\dagger}_{}\phi)^{2}_{}+\frac{1}{2}\lambda_1^{}\sigma^2_{}\chi^2_{}\nonumber\\
&&+\lambda_3^{}\sigma^2_{}\phi^{\dagger}_{}\phi+\lambda_3^{}\chi^2_{}\phi^{\dagger}_{}\phi\,,
\end{eqnarray}
The non-zero VEV of the singlet scalar $\xi$ gives masses to the
right-handed neutrinos, while the VEV of the SM Higgs doublet
$\phi$ breaks the electroweak symmetry and give Dirac masses to
all the fermions. Together, they give rise to the tiny masses of left-handed neutrinos through the seesaw mechanism. Minimising
the potential in Eq.~\ref{potential_f}, one finds non-zero VEVs of
these two scalars \cite{us},
%\begin{subequations}
\begin{eqnarray}
u&=&\sqrt{2}\langle\sigma\rangle =\sqrt{
\frac{\lambda_2^{}(\mu_1^2+\mu_3^2)-\lambda_3^{}\mu_2^2}{\lambda_1^{}\lambda_2^{}-\lambda_3^2}}\,,\\
v&=&\sqrt{2}\langle\phi\rangle =\sqrt{
\frac{\lambda_1^{}\mu_2^2-\lambda_3^{}(\mu_1^2+\mu_3^2)}{\lambda_1^{}\lambda_2^{}-\lambda_3^2}}\,.
\end{eqnarray}
%\end{subequations}
We can redefine the scalar $\sigma$ and $\phi$ as,
\begin{eqnarray}
\sigma=\frac{1}{\sqrt{2}}(u+\Phi)\quad\textrm{and}\quad
\phi=\frac{1}{\sqrt{2}}\left[
\begin{array}{c}
v+H\\
[3mm] 0 \end{array} \right]\,.
\end{eqnarray}
The VEV $v=246$ GeV has been determined from the electroweak
symmetry breaking.  We will argue in the following that, in order
to establish the pseudo-Majoron as an alternative candidate for
cold dark matter, the VEV $u$ should be near grand unified theory
(GUT) scale ($u\sim 10^{16}$ GeV). $\chi$ has no mixing with
$\Phi$ and $H$ because of its zero VEV. As a consequence of the
huge hierarchy between $u$ and $v$, the mixing between $\Phi$ and
$H$ is extremely small so that $H$ can be identical to the SM
Higgs boson. The mass terms of the physical Higgs scalars can be
written as,
\begin{eqnarray}
\mathcal{L}_{m}^{}\supset
-\frac{1}{2}m_{\Phi}^{2}\Phi^2_{}-\frac{1}{2}m_{H}^{2}H^2_{}-\frac{1}{2}m_\chi^2\chi^2_{}
\end{eqnarray}
with
\begin{eqnarray}
m_{\Phi}^{2}\simeq 2\lambda_1^{}u^2_{}\,,~~ m_{H}^{2}\simeq
2(\lambda_2^{}-\frac{\lambda_3^2}{\lambda_1^{}})v^2_{}\,,~~
m_\chi^2=2\mu_3^2\,.
\end{eqnarray}

In order to establish the pseudo-Majoron, $\chi$, as a candidate for cold dark matter, its lifetime should be long enough. The pseudo-Majoron couples to the right-handed neutrinos (see Eq.~\ref{yukawa}),
\begin{eqnarray}
\label{yukawa3} \mathcal{L}\supset
-\frac{i}{2\sqrt{2}}h\chi\bar{N}_R^{c}N_R^{}+\textrm{H.c.}\,.
\end{eqnarray}
For $m_{\chi}\ll M_N^{}$, the pseudo-Majoron will decay into
the SM particles through the virtual right-handed neutrinos.
Conveniently, we can integrate out the heavy right-handed neutrinos
to derive the effective couplings of the pseudo-Majoron to the
left-handed neutrinos,
\begin{eqnarray}
\label{yukawa4} \mathcal{L}_{eff}^{}=
i\frac{m_\nu^{}}{2u}\chi\bar{\nu}_L^{}\nu_L^c\left(1+\frac{H}{v}\right)^2_{}+\textrm{H.c.}\,.
\end{eqnarray}
The effective couplings of the pseudo-Majoron to the left-handed neutrinos are highly suppressed by the neutrino mass in the numerator and by the VEV $u$ in the denominator. It was shown in Ref.~\cite{us} that for the VEV $u\sim 10^{16}$ GeV, the decay width of the pseudo-Majoron into a pair of SM neutrinos is highly suppressed and the lifetime can be very long. As a successful dark matter candidate, its relic density should be consistent with the cosmological observations. It has been shown in earlier works \cite{dm1,dm2,dm3} that a stable SM-singlet scalar with a quartic coupling to the SM Higgs doublet can serve as the dark matter because it contributes a desired relic density through the annihilations into the SM particles. In the present model, the pseudo-Majoron $\chi$ also has a quartic coupling with the SM Higgs,
\begin{eqnarray}
\label{quartic}
V\supset\lambda_3^{}\chi^2_{}\phi^\dagger_{}\phi\Rightarrow
\lambda_3^{} v \chi^2_{} H+\frac{1}{2}\lambda_3^{}
\chi^2_{}H^2_{}\,.
\end{eqnarray}
This implies that the pseudo-Majoron $\chi$ with its very long
lifetime can play the role of the dark matter\footnote{It is important to mention that for a sizable $\lambda_3$, the pattern of present symmetry breaking i.e. $u\sim 10^{16}~{\rm GeV} >>v\sim 264$ GeV, requires a fine tuning between $\lambda_3 u^2$ and $\mu_2^2$ so that $\lambda_3 u^2-\mu_2^2$ can be of the order of $v^2$ \cite{us}.}.

\begin{figure}[t]
\begin{center}
\epsfig{file=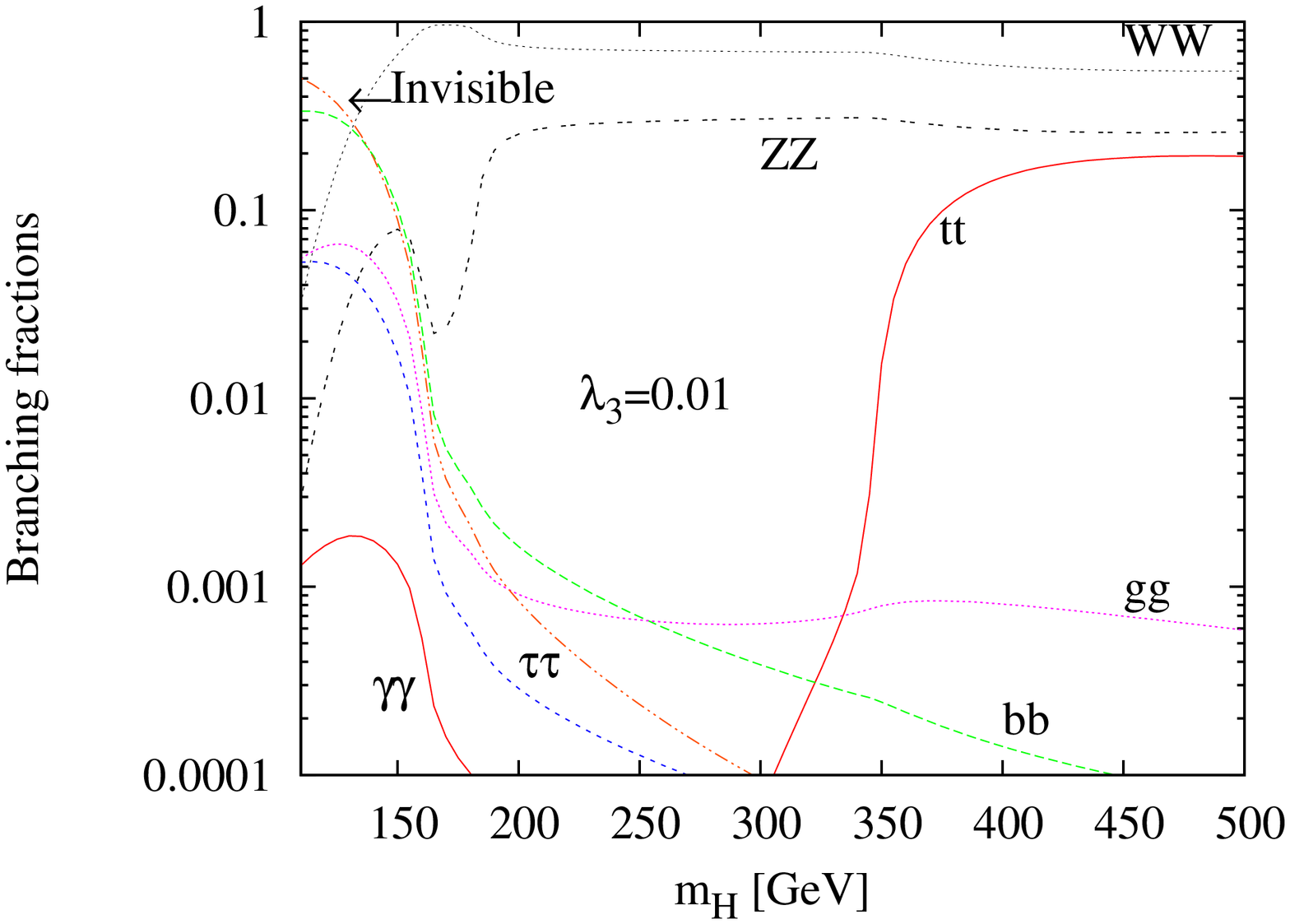,width=6cm,height=5cm,angle=270}
\epsfig{file=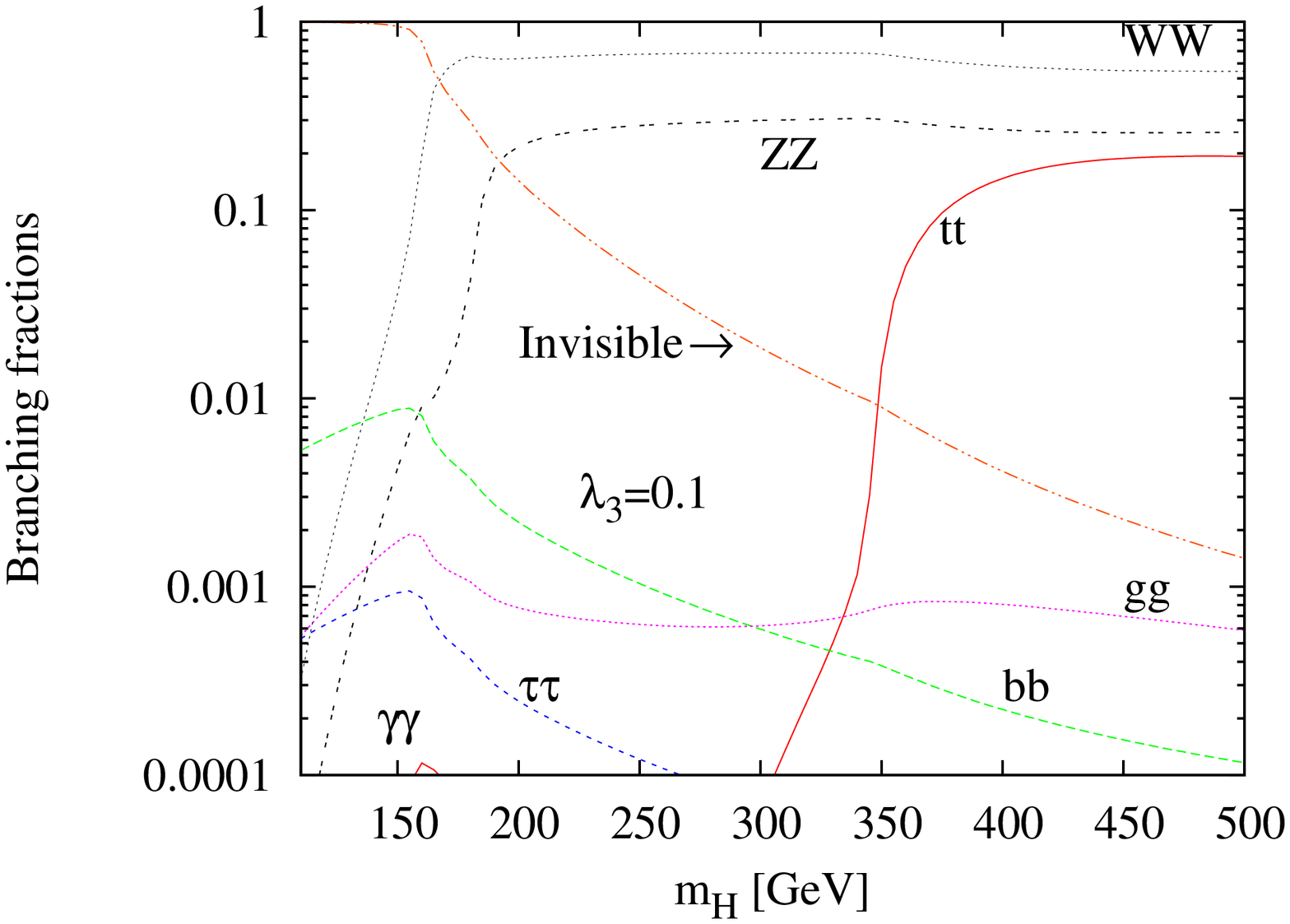,width=6cm,height=5cm,angle=270}
\epsfig{file=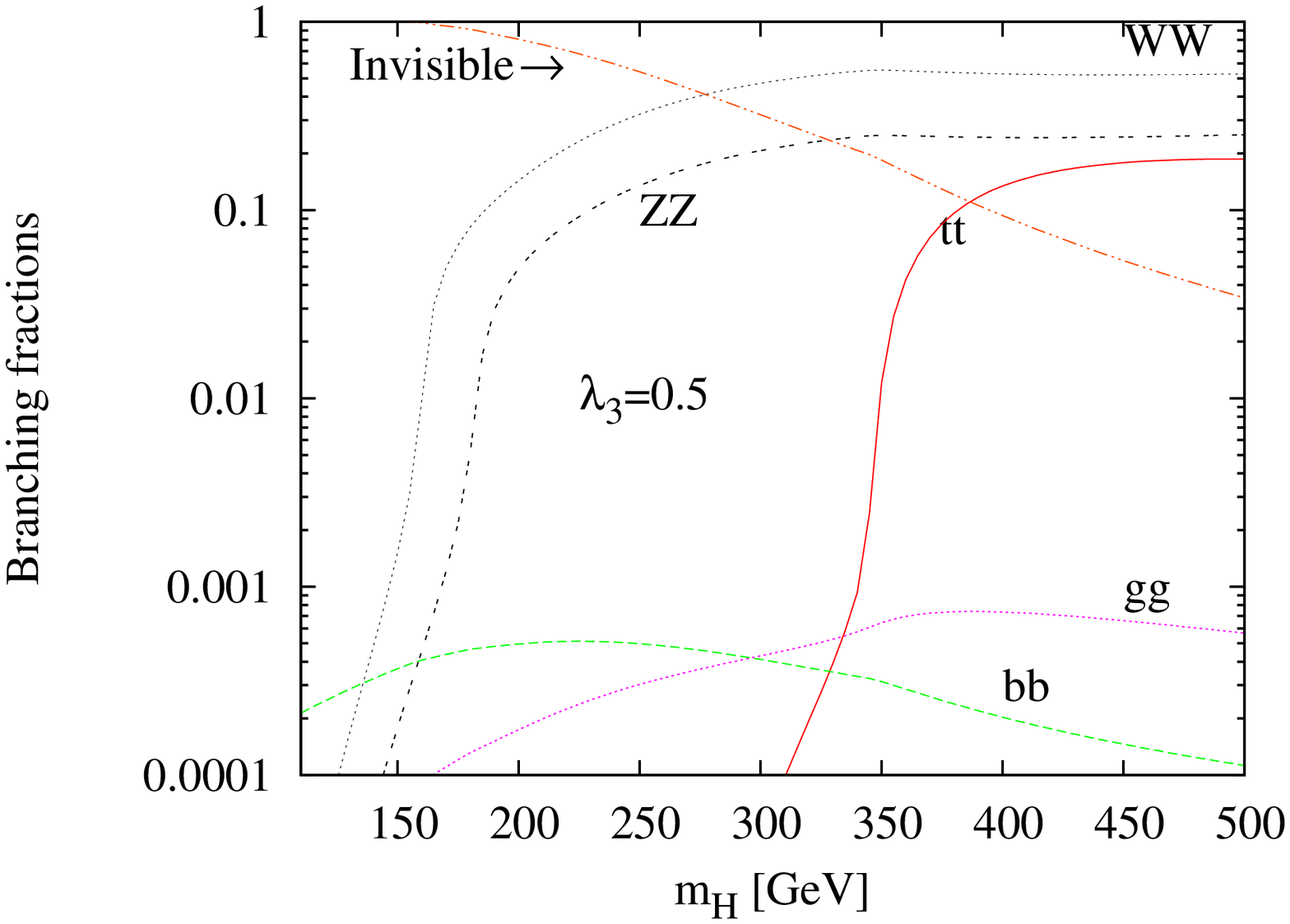,width=6cm,height=5cm,angle=270}
\end{center}
\caption{Decay branching fraction of Higgs boson as a function of Higgs mass for three different values of $\lambda_3$.}
\label{fig:higgsBR}
\end{figure}

The coupling of the pseudo-Majoron $\chi$ to the SM Higgs $H$ in Eq.~\ref{quartic} not only determines the dark matter relic density but also opens a new decay mode for the Higgs boson. When it is kinematically possible, Higgs can mostly decay into a pair of pseudo-Majorons and gives rise to missing energy signature at the LHC. Decay width of the Higgs boson into a pair of pseudo-Majorons is given by,
\begin{equation}
\Gamma(H\to \chi \chi)=\frac{1}{8\pi m_H^2}\left( \lambda_3 v \right)^2 \left({m_H^2-4m_\chi^2}\right)^\frac{1}{2}~,
\label{eq:higgsdecay}
\end{equation}
where, $m_H$ is the mass of Higgs and $m_\chi$ is the mass of the pseudo-Majoron. We have used the code HDECAY \cite{HDECAY} to calculate the decay widths and the branching ratios of the Higgs boson ($H$). In Fig.~\ref{fig:higgsBR}, we present the decay branching fraction of Higgs boson as a function of Higgs mass for three different values of $\lambda_3$ ($\lambda_3=0.01,~0.1~{\rm and}~0.5$). From Fig.~\ref{fig:higgsBR}, one can notice that for low Higgs mass ($m_H<150$ GeV), invisible decay mode dominates over the $b \bar b$ mode for $\lambda_3=0.01$. For larger $\lambda_3=0.1~(0.5)$, invisible decay mode remains the dominant one upto Higgs mass 160 (275) GeV. The invisible decay of Higgs boson significantly modifies the collider searches for low mass Higgs boson. Due to less SM background, diphoton is the most promising channel for the discovery of low mass Higgs boson. However, as can be seen from Fig.~\ref{fig:higgsBR}, in presence of large invisible decay width of the Higgs boson, the diphoton branching ratio gets extremely suppressed, thus jeopardising discovery via this channel at the LHC.
%------------------------------------------------------------------
%-------------------------------------------------------------------

\begin{figure}[t]
\begin{center}
\epsfig{file=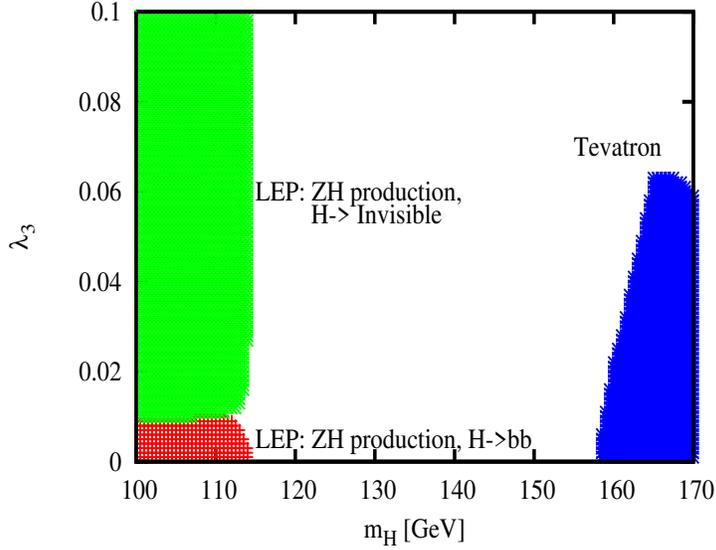,width=8cm,height=10cm,angle=270}
\end{center}
\caption{95\% C.L. excluded regions in the $m_H-\lambda_3$ plane. The colours indicates the channel for highest statistical sensitivity. Red: LEP $e^+e^-\to ZH,~H\to b \bar b$, Green: LEP $e^+e^-\to ZH,~Z\to$ invisible, Blue: Tevatron $p \bar p \to H+X$.}
\label{fig:bound}
\end{figure}

Results from the LEP and the Tevatron experiment can also be translated to restrict the allowed region in the $m_H-\lambda_3$ space. Non-observability of any Higgs boson signal at the LEP experiment puts bound on the Higgs mass and its invisible decay branching fraction. In the framework of the present model, Higgs invisible branching fraction is a function of both $\lambda_3$ and $m_H$. We have used the package {\bf HiggsBounds-2.0.0} \cite{hb1,hb2} which tests theoretical prediction of models with an arbitrary Higgs sector against the exclusion bounds obtain from the LEP and Tevatron. In Fig.~\ref{fig:bound}, we have presented 95\% C.L. excluded regions in the $m_H-\lambda_3$ plane from the direct detection experiments. At the LEP, the Higgs boson is expected to be produced mainly via the Higgs-strahlung process: $e^+e^-\to ZH$. For low values of $\lambda_3$, invisible decay mode of the Higgs boson is highly suppressed. Therefore, for low $\lambda_3 (<0.01)$, Higgs mass is excluded upto 114 GeV at 95\% C.L. (see Fig.~\ref{fig:bound}) from LEP Higgs search in $b \bar b$ channel \cite{hboundbb}. In view of the fact that Higgs boson can decay invisibly, four LEP collaborations performed searches for acoplanar jets ($H\to invisible$)($Z\to q \bar q$) \cite{hboundinvi}. For larger values of $\lambda_3$, invisible branching fraction dominates over the $b \bar b$ mode. Fig.~\ref{fig:bound} shows that in this part of the parameter space, Higgs mass is also excluded upto 114 GeV from the LEP search for the invisible Higgs boson. However, in the region $\lambda_3 \sim 0.01$, $b \bar b$ and invisible branching fractions become comparable and the LEP exclusion limit of Higgs mass is slightly lower. Direct searches for the standard model (SM) Higgs boson at the Tevatron exclude a new and larger region at high mass between $158 < m_H < 175$ GeV at 95\% C.L. \cite{hboundtev}. However, in presence of a invisible decay mode, the Tevatron exclusion limit on the Higgs mass should be modified. One should in particular note that for $\lambda_3>0.06$ Tevatron bound on the Higgs mass does not exist (see Fig.~\ref{fig:bound}).

%------------------------------------------------------------------
%-------------------------------------------------------------------
\section{Collider Signature}

It has been already demonstrated that in the frame work of the present pseudo-Majoron model, a low mass Higgs dominantly decays into a pair of pseudo-Majorons which are stable and elusive at the detector. To extract the signature of such an invisible Higgs at the LHC, we note that the most dominant associated production channels, namely, gluon-gluon fusion ($gg\to H+{\rm jet}$) and vector boson fusion ($qq\to qqV^*V^*\to qqH$) channel, result into purely hadronic final state together with missing transverse energy. QCD backgrounds become quite serious for them, although some suggestions have been made in the context of vector boson fusion \cite{Datta:2001hv,Eboli:2000ze}. In this work, we consider the production of the Higgs in association with a $W$ or $Z$-boson, where, again from the standpoint of background suppression, we have considered the leptonic decay modes of $W$ and $Z$-boson.

The production of $WH$ is mediated by a virtual $W$-boson in the s-channel. Leptonic decay of $W$-boson ($W\to l \nu$) and invisible decay of the Higgs ($H\to \chi\chi$) gives rise to single hard lepton $+$ missing transverse energy.
\begin{equation}
q \bar q^{\prime} \to W^{*} \to WH \to (l \nu)(\chi \chi)\nonumber
\end{equation}
\begin{itemize}
\item The irreducible SM background to the $1l+E_T\!\!\!\!\!\!/~~$ signal arises from $WZ$ production where $W$ decays leptonically and $Z$ decays invisibly. However, this background is suppressed by the invisible branching ratio of the $Z$-boson.
\item The most significant background to the single lepton $+~E_T\!\!\!\!\!\!/~~$ signal arises from the single $W$ production followed by the leptonic decay of the $W$-boson. The single $W$-production cross-section at the LHC is huge. However, one can define the transverse mass: $M_T=\sqrt{2p_T^l p_T\!\!\!\!\!/~[1-cos\phi(\vec p_T^l,\vec p_T\!\!\!\!\!\!/~)]}$ and demand $M_T > 100$ GeV to remove the background coming from real $W$ production. However, this cut cannot suppress the background coming from virtual $W^*$ production which is order of magnitude larger than the signal. Therefore, detecting the signature of invisible Higgs boson in $WH$ channel is extremely challenging. In our analysis, we do not consider $1l+E_T\!\!\!\!\!\!/~~$ as signature of invisibly decaying Higgs.
\end{itemize}

We find it more convenient to use the associated production of $ZH$ followed by the leptonic decay of $Z$-boson and invisible decay of Higgs gives rise to two unlike sign, same flavour leptons $+$ missing transverse energy signature at the LHC.
\begin{equation}
q \bar q \to Z^{*} \to ZH \to (l \bar l)(\chi \chi)\nonumber
\end{equation}
The signal is characterized by a peak in the dilepton ($M_{ll}$) invariant mass distribution at the Z-boson mass ($m_Z$). The dominant backgrounds for the dilepton $+E_T\!\!\!\!\!\!/~~$ signal are listed below:
\begin{itemize}
\item The irreducible background to the signal comes from the pair production of $Z$-boson ($q\bar q\to ZZ$) followed by the leptonic decay of one $Z$ and invisible ($\nu \bar \nu$) decay of other. Although the mass of $Z$-boson and Higgs are different, all other kinematic properties of $ZH$ signal and $ZZ$ background are identical. In case of both signal and background leptons come from the decay of $Z$-boson, dilepton invariant mass distribution cannot be used to separate the signal. On the other hand, the missing $E_T$  for the signal arises from the invisible decay of Higgs and for background, invisible decay of $Z$-boson gives rise to the missing transverse energy. However, in an environment like LHC where the longitudinal boost of the center-of-mass frame is unknown, it is not possible to determine the mass of the particle decaying invisibly. Therefore, it is not possible to suppress this background without reducing the signal.
\item Pair production of $W$-boson followed by the leptonic decay of both the $W$ also contribute to the dilepton $+~E_T\!\!\!\!\!\!/~~$ background. Since the signal dileptons are characterised by a peak (at $m_Z$) in the dilepton invariant mass distribution, we can suppress this background contribution significantly by putting a cut on the dilepton invariant mass ($M_{ll}$).
\item The production of top anti-top pairs may also give rise to $2l+E_T\!\!\!\!\!\!/~~$ when both top and anti-top decays leptonically and two b-jets fall out side detector coverage.
\item Single $Z$-boson production can also contribute significantly to the background. Here, missing transverse momentum arises from the high energy ISR jets which get lost in the beam pipe, and also from the jet energy mismeasurment.
\end{itemize}

In this analysis, we have generated the signal and SM background events with PYTHIA 6.421~\cite{PYTHIA}. We have used the leading order CTEQ6L1~\cite{CTEQ} parton distribution functions. The QCD factorization and renormalization scales are kept fixed at $\sqrt s$.

\subsection{Identifying jets, leptons e.t.c.}
In our analysis, we have introduced a set of basic selection criteria to identify leptons,
jets and missing transverse energy. The object selection is described in brief in the following.\\\\
\noindent
{\bf Lepton selection:}
\begin{itemize}
 \item $p_T>10 \gev$ and $|\eta_{\ell}|<$ 2.5, where $p_T$ is
   the transverse momentum and $\eta_{\ell}$ is the pseudorapidity
   of the lepton (electron or muon).
\item {\bf Lepton-lepton separation:}  ${\Delta R}_{\ell\ell} \geq $ 0.2,
  where $\Delta R = \sqrt {(\Delta \eta)^2 + (\Delta \phi)^2}$ is the
  separation in the pseudorapidity--azimuthal angle plane.
\item {\bf Lepton-jet separation:} $\Delta R_{\ell j} \geq 0.4$ for all jets
  with $E_T >$ 20 GeV.
\item The total energy deposit from all {\it hadronic activity} within a cone
  of $\Delta R \leq 0.2$ around the lepton axis should be $\leq$ 10 GeV.
\end{itemize}

\noindent
{\bf Jet selection:}

\begin{itemize}
\item
Jets are formed with the help of {\tt PYCELL}, the inbuilt cluster routine in
\PYTHIA.  The minimum $E_{T}$ of a jet is taken to be $20\gev$, and
we also require  $|\eta_j|<$ 2.5.
\end{itemize}
\noindent
{\bf Missing transverse energy ($E_T\!\!\!\!\!\!/~~$):}\\
The missing transverse energy in an event is calculated using calorimeter cell energy and
the momentum of the reconstructed muons in the muon spectrometer.
In our analysis, we have used the following definition for the missing transverse energy:

\begin{equation}
E_{T}\!\!\!\!\!\!/~~=\sqrt{(\sum p_x)^2+(\sum p_y)^2},
\label{met}
\end{equation}
where, the sum goes over all the isolated electrons, muons, the jets as well as the `unclustered' energy deposits.

We have approximated the detector resolution effects by smearing the
energies (transverse momenta) with Gaussian
functions.
%~\cite{TDR,Mellado}.
The different contributions to the
resolution error have been added in quadrature.

\begin{itemize}
\item {\bf Electron energy resolution:}
\be
\frac{\sigma(E)}{E} = \frac{a}{\sqrt{E}} \oplus b \oplus
                      \frac{c}{E},
\ee
where
\be
(a, b, c) = \left\{ \begin{array}{lcl}
                   (0.030\gev^{1/2}, \, 0.005, \, 0.2\gev), & \hspace{1em} &
                                                   |\eta| < 1.5, \\
                    (0.055\gev^{1/2}, \, 0.005, \, 0.6\gev), & \hspace{1em} &
                           1.5 < |\eta| < 2.5.
            \end{array}
            \right.
\ee
\item {\bf Muon $p_T$ resolution:}
\be
\frac{\sigma(p_T)}{p_T} = \left\{ \begin{array}{lcl}
                       a , & \hspace{1em} &  p_{T} < 100\gev, \\
                       \displaystyle
                         a + b \, \log \frac{p_T}{100\gev} , & &
             p_{T}>100\gev,
            \end{array}
            \right.
\ee
with
\be
(a, b) = \left\{ \begin{array}{lcl}
                       (0.008, \, 0.037), & \hspace{1em} & |\eta| < 1.5,  \\
                       (0.020, \, 0.050), & \hspace{1em} & 1.5 < |\eta| <
                                                           2.5. \\
            \end{array}
            \right.
\ee
\item {\bf Jet energy resolution:}
\be
\frac{\sigma(E_T)}{E_T} = \frac{a}{\sqrt{E_T}},
\ee
with
$ a= 0.5\gev^{1/2}$, the default value used in {\tt PYCELL}.
\end{itemize}
  %------------------------------------------------------------------

\begin{table}[h]

\begin{center}

\begin{tabular}{||c|c|c|c|c|c||}
\hline \hline
\multicolumn{6}{||c||}{Cross-section in fb} \\\hline\hline
\multicolumn{6}{||c||}{LHC with $\sqrt s=14$ TeV} \\\hline\hline
        & \multicolumn{5}{|c||}{Background} \\\cline{2-6}
Process & Cut-1 & Cut-1 & Cut-1 &  Cut-1 & Cut-1 \\
        &       & $+E_T\!\!\!\!\!\!\!/~>100$ GeV & $+E_T\!\!\!\!\!\!\!/~>150$ GeV & $+E_T\!\!\!\!\!\!\!/~>170$ GeV & $+E_T\!\!\!\!\!\!\!/~>200$ GeV \\\hline\hline
$ZZ$    & 104.95 & 12.58 & 4.28 & 2.98 &1.75\\
$W^+W^-$& 156.85 & 0.76 & 0.15 & $<$0.15 & $<$0.15 \\
$tt$    & 76.8 & 5.32 & 0.42 & $<$0.2 & $<$0.2 \\
$Z$     & 7.8$\times 10^5$ & 70.30 & $<$0.35 & $<$0.35&$<$0.35\\\hline\hline
$m_H$ &\multicolumn{5}{|c||}{Signal} \\\hline\hline
$120$ GeV & 17.81 &        5.05 &       2.18 &        1.56 &       0.99 \\
$160$ GeV &  7.08 &        2.92 &        1.49 &        1.16 &       0.78 \\\hline\hline
\multicolumn{6}{||c||}{LHC with $\sqrt s=10$ TeV} \\\hline\hline
Process & \multicolumn{5}{|c||}{Background} \\\hline\hline
%Process & Acceptance & Acceptance & Acceptance &  Acceptance & Acceptance \\
%        &   cuts   & $+E_T\!\!\!\!\!\!\!/~>100$ GeV & $+E_T\!\!\!\!\!\!\!/~>150$ GeV & $+E_T\!\!\!\!\!\!\!/~>170$ GeV & $+E_T\!\!\!\!\!\!\!/~>200$ GeV \\\hline\hline
$ZZ$    & 77.84 & 8.89 & 2.83 & 1.70 & 0.83 \\
$W^+W^-$& 120.62 & 0.29 & $<$0.09 & $<$0.09 & $<$0.09 \\
$tt$    &  16.71 & 1.39 & $<$0.23 & $<$0.23 & $<$0.23 \\
$Z$     & 6.1$\times 10^5$ & 12.49 & $<$0.2 & $<$0.2&$<$0.2\\\hline\hline
%$Z$     & - & - & - & - & - \\\hline\hline
$m_H$ &\multicolumn{5}{|c||}{Signal} \\\hline\hline
$120$ GeV & 12.14 &        3.21 &       1.37 &        1.02 &       0.66 \\
$160$ GeV &  4.81 &        1.83 &        0.91 &        0.69 &       0.46 \\\hline\hline
\end{tabular}

\end{center}

\caption{Signal (for $m_H=120~{\rm and}~160$ GeV) and the SM background cross-sections after {\bf Cut-1} and {\bf Cut-1}$+E_T\!\!\!\!\!\!/~~>$ 100, 150, 170 and 200 GeV at the LHC with $\sqrt s=$ 10 and 14 TeV.}

\label{table:SB}
\end{table}

%--------------------------------------------------------------------

\subsection{Event Selection}
After introducing the basic object selection criteria and discussing about the signal and background characteristics, we are now equipped enough to introduce an additional set of event selection criteria which will enhance the signal to background ratio.

\begin{figure}[t]
\begin{center}
\epsfig{file=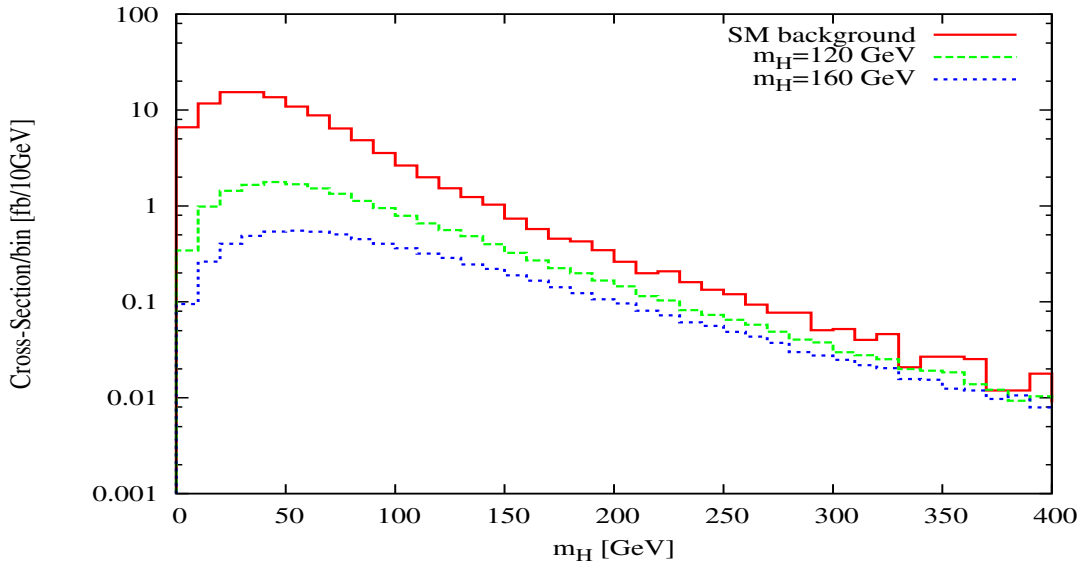,width=8cm,height=15cm,angle=270}
\end{center}
\caption{Missing $E_T$ distributions for signal (with $m_H=$120 and 160 GeV) and dominant $ZZ$ background after {\bf Cut-1} at the LHC with $\sqrt s=14$ TeV. In this plot, we have assumed that Higgs decays invisibly with 100\% branching fraction.}
\label{fig:met}
\end{figure}

\begin{itemize}
\item We demand exactly one pair of leptons with same flavour and opposite charge and $p_T>20$ GeV. We reject events with any additional lepton with $p_T > 20$ GeV.
\item We have considered hadronically quiet opposite sign dilepton events {\em i.e.}, we veto events with one or more central ($|\eta|<2.5$) jets having $p_T>20$ GeV.
\item Since the signal leptons result from the decay of $Z$-boson, we reject events with $|M_{ll}-m_Z|>10$ GeV. This cut will reduce the $W^+W^-$ and $t\bar t$ background significantly.
\end{itemize}
We collectively refer to the basic isolation cuts (described in section 3.1), the abovementioned $p_T$ and dilepton invariant mass cuts as {\bf Cut-1}. In Fig.~\ref{fig:met}, we have presented the missing transverse momentum distributions for signal (with $m_H=120~{\rm and}~160$ GeV) and dominant $ZZ$ background after {\bf Cut-1} at the LHC with $\sqrt s=14$ TeV. In Fig.~\ref{fig:met}, we have assumed that Higgs decays invisibly with 100\% branching fraction. Fig.~\ref{fig:met} shows that differential cross-section decreases rapidly with $E_T\!\!\!\!\!\!/~~$, however, there is a clear enhancement of signal to background ratio in the high $E_T\!\!\!\!\!\!/~~$ region. This enhancement is more pronounced for larger Higgs mass.  In Table~\ref{table:SB}, we have presented the signal (for $m_H=120~{\rm and}~160$ GeV) and different SM background cross-sections after {\bf Cut-1} and {\bf Cut-1} plus four different $E_T\!\!\!\!\!\!/~~$ ($E_T\!\!\!\!\!\!/~~>$100, 150, 170 and 200 GeV) cuts at the LHC with $\sqrt s=$10 and 14 TeV\footnote{We have checked that the luminosity planned for $\sqrt s=7$ TeV is not sufficient to yield signal of the type suggested by us. Although, the current plan is to upgrade directly to $\sqrt s=14$ TeV, we are nonetheless presenting results for $\sqrt s=10$ TeV and the luminosity required by us there, in case the occasion arises in the future.}. In view of the cross-sections in Table~\ref{table:SB}, we choose {\bf Cut-1} $+~E_T\!\!\!\!\!\!/~~>150$ GeV as our final selection criteria.

%------------------------------------------------------------------
%-------------------------------------------------------------------

\subsection{Discovery reach}
\begin{figure}[t]
\begin{center}
\epsfig{file=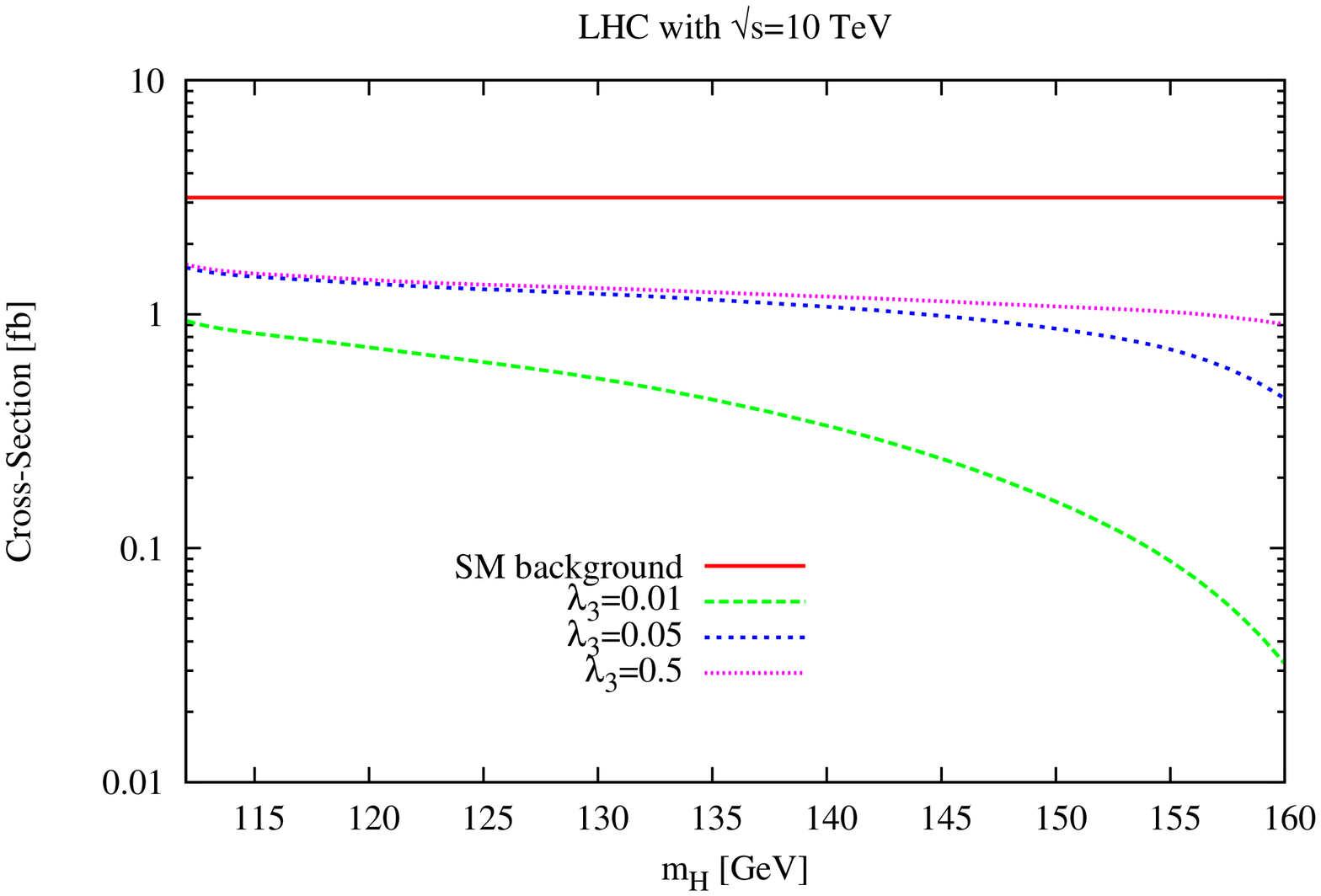,width=8cm,height=8cm,angle=270}
\epsfig{file=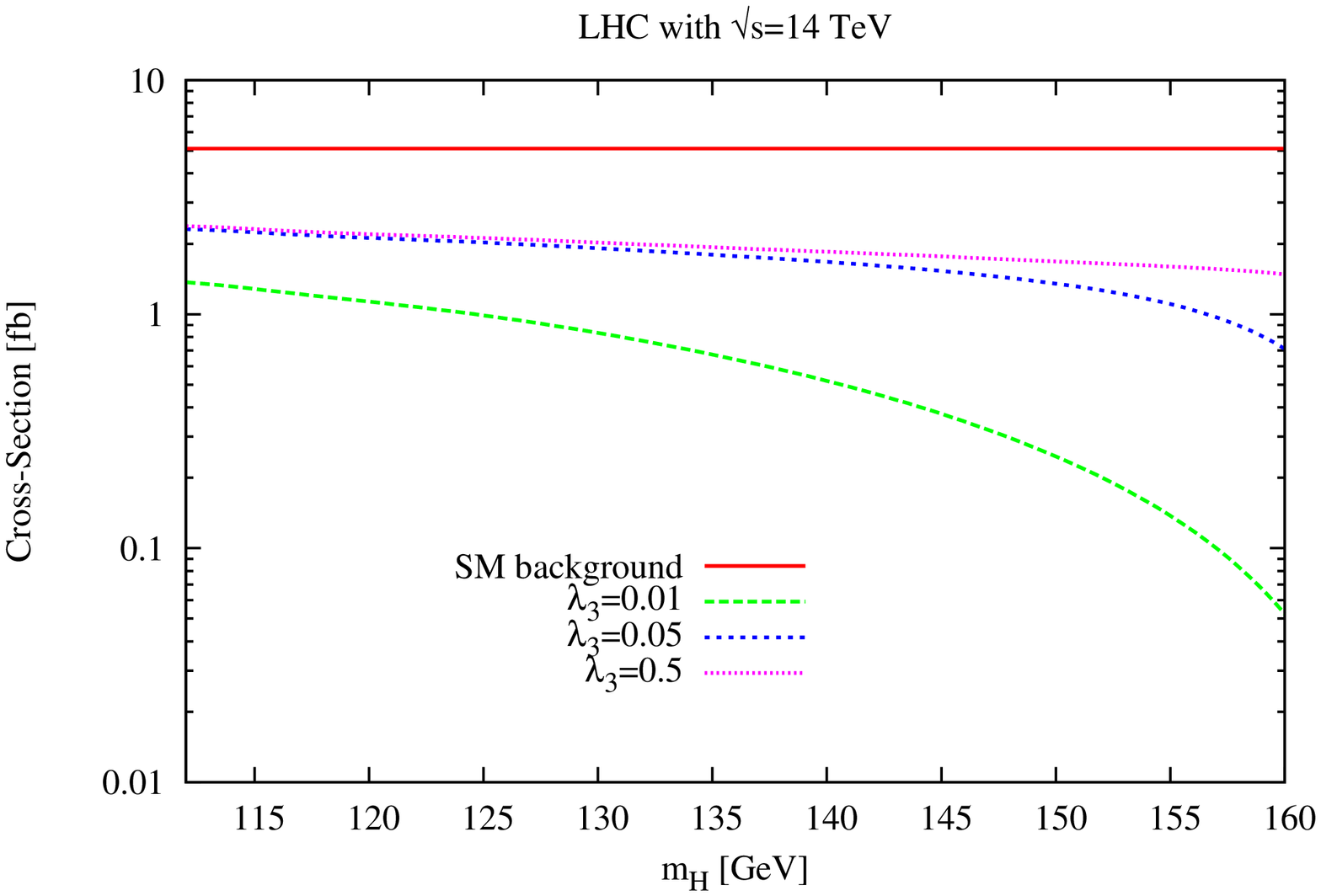,width=8cm,height=8cm,angle=270}
\end{center}
\caption{Signal cross-section (after selection cuts) for three different values of $\lambda_3$ as a function of Higgs mass at the LHC with centre-of-mass energy 14 TeV (right) and 10 TeV (left). The total SM background cross-section is also presented in the figure.}
\label{fig:cross}
\end{figure}

With the criteria listed already, we obtained the discovery
potential of invisible Higgs (in the frame work of present
pseudo-Majoron model) at the LHC, with centre-of-mass energy 10
and 14 TeV. The production cross-section of $ZH$ depends only on
the mass of the Higgs boson ($m_H$). We choose $m_H$ as one of the
scan parameter. The invisible decay branching fraction of Higgs
depends on both $m_H$ and the quartic coupling $\lambda_3$ (see
Fig.~\ref{fig:higgsBR}). Therefore, the resulting dilepton
$+~E_T\!\!\!\!\!\!/~~$ signal cross-section depends on both $m_H$
and $\lambda_3$. We have chose $\lambda_3$ as the second scan
parameter. To show the variation of signal with $m_H$ and
$\lambda_3$, in Fig.~\ref{fig:cross}, we have presented the signal
cross-section as a function of Higgs mass for three different
values of $\lambda_3$ (=0.5, 0.05 and 0.01). For large values of
$\lambda_3$, the invisible decay mode of the Higgs is
overwhelmingly dominant over the other decay modes. Therefore, the
variation of signal cross-section with $m_H$ for large $\lambda_3$
is not very significant. For small $\lambda_3$, the invisible
Higgs decay mode still dominant in the low $m_H$ region however,
for $m_H>150$ GeV, $H\to WW^*~{\rm and}~ZZ^*$ decay modes open up
and the invisible decay branching fraction decreases sharply (see
Fig.~\ref{fig:higgsBR}). As a result, in Fig.~\ref{fig:cross}, we
observe a sharp fall in the signal cross-section above $m_H=150$
GeV for $\lambda_3=0.01$.
%------------------------------------------------------------------
%-------------------------------------------------------------------

We define the signal to be observable for an integrated luminosity ${\cal L}$ if \cite{bt1,bt2},
\begin{itemize}
\item
\begin{equation}
\frac{N_{S}}{\sqrt{N_B+N_S}} \ge 5 ~~~~~~ {\rm for}~~~~~ 0< N_B \le 5 N_S,
\end{equation}
where, $N_{S(B)}=\sigma_{S(B)} {\cal L}$, is the number of signal (background) events for an integrated
luminosity ${\cal L}$.
\item For zero background event, we treat the signal as decisive if there are at
least five signal events.
\item In order to establish the discovery of a small signal (which could be
statistically significant i.e. $N_S/\sqrt{N_B}\ge 5$) on top of a large background, we need to know the
background with high precision. However, such precise determination of the SM background is beyond
the scope of this present article. Therefore, we impose the requirement $N_B\le 5 N_S$ to avoid such
possibilities.
\end{itemize}

\begin{figure}[t]
\begin{center}
\epsfig{file=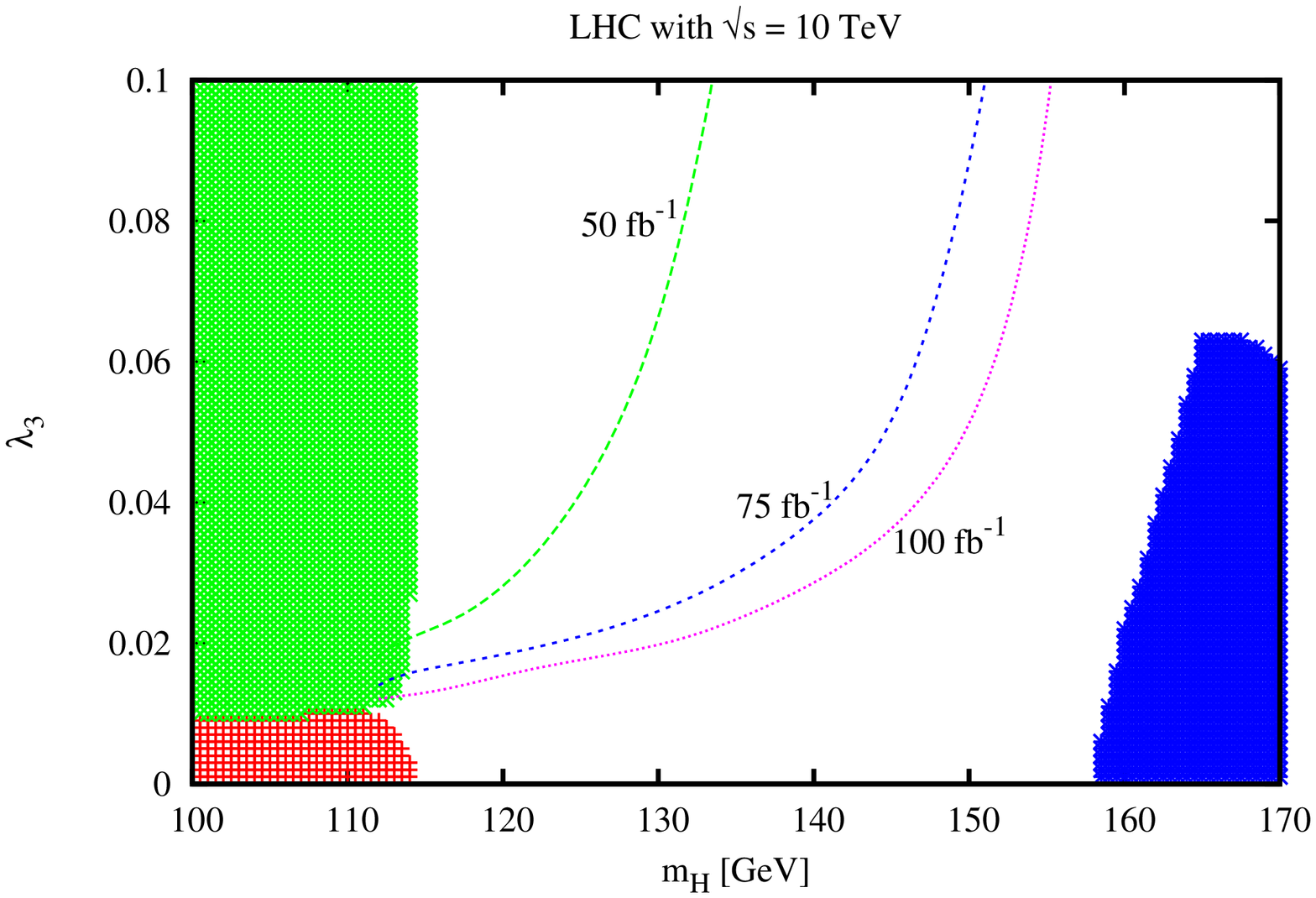,width=8cm,height=8cm,angle=270}
\epsfig{file=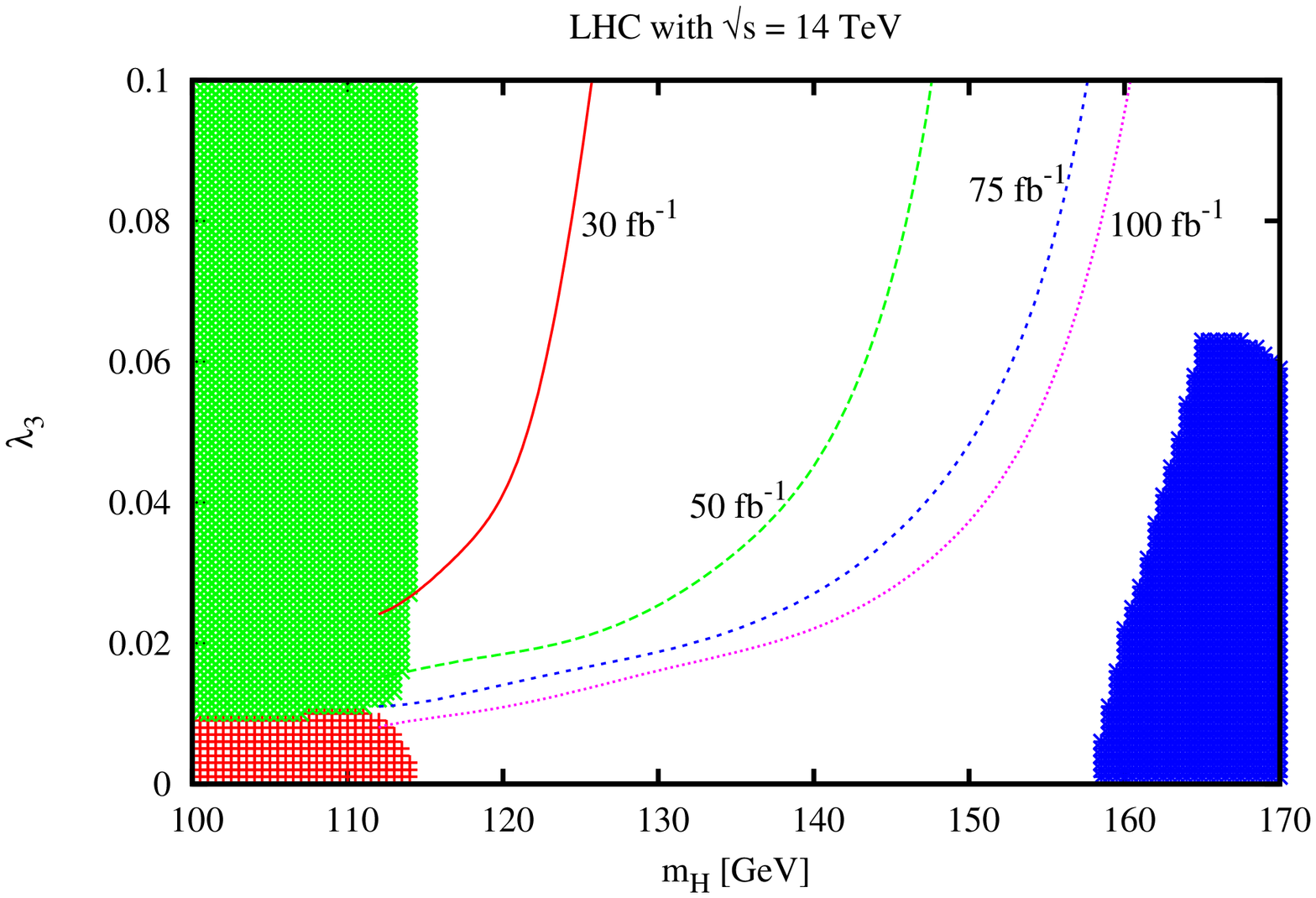,width=8cm,height=8cm,angle=270}
\end{center}
\caption{Discovery reach of invisibly decaying Higgs boson in the frame work of pseudo-Majoron model at the LHC with center-of-mass energy 10 TeV (right) and 14 TeV (left). Different values of the integrated luminosity are assumed. Each line corresponds to $5\sigma$ discovery contour in the $m_H-\lambda_3$ plane. The shaded regions represent the experimentally excluded parts of the $m_H-\lambda_3$ space (see Fig.~\ref{fig:bound}).}
\label{fig:lumi}
\end{figure}

In Fig.~\ref{fig:lumi}, we have presented the discovery potential of invisible Higgs (in the framework of the present pseudo-Majoron) at the LHC with center-of-mass energy 10 TeV (right) and 14 TeV (left). We have assumed different
values of integrated luminosity ranging from $30~{\rm fb}^{-1}$ to $100~{\rm fb}^{-1}$. Fig.~\ref{fig:lumi} shows the 5$\sigma$ discovery contours in the $m_H$-$\lambda_3$ plane and the lines refer to the different integrated luminosities. With $100~{\rm fb}^{-1}$ of integrated luminosity and 14 (10) TeV center-of-mass energy of the LHC, the invisible decay of the Higgs boson in the large $\lambda_3$ ($>0.1$) region can be probed upto $m_H=160~(155)$ GeV. For $\lambda_3\leq 0.02$, the LHC with $100~{\rm fb}^{-1}$ of luminosity and $\sqrt s=14~(10)$ TeV can discover the invisible Higgs boson upto $m_H=140~(135)$ GeV.

%------------------------------------------------------------------
%-------------------------------------------------------------------

\section{Conclusion}
We have investigated the collider phenomenology of a particular variant of the singlet Majoron model where the pseudo-Majoron is a potential dark matter candidate. This is achieved through the soft breaking of global lepton number. The quartic coupling of the pseudo-Majoron with the SM Higgs allows the SM Higgs to decay in to a pair of pseudo-Majorons  whenever it is kinematically possible. We found that for low Higgs mass, the decay of the Higgs into a pair of pseudo-Majorons always dominates over the SM decay modes. We consider the production of the Higgs boson in association with a $Z$-boson followed by the leptonic decay of the $Z$. Our signal consists of two hard isolated lepton  in presence of missing $E_T$. The pair production of the $Z$-boson is the only irreducible SM background. We have define a set of event selection criteria to enhance the signal to background ratio. With this event selection criteria, we found that at the LHC with $\sqrt s=14$ TeV and $100~{\rm fb^{-1}}$ integrated luminosity, the invisible decay of the Higgs can be probed upto $m_H=160$ GeV for values close to $0.1$ for the Higgs pseudo-Majoron quartic coupling.

\section*{Acknowledgments} 

KG and BM was partially supported by funding available from the
Department of Atomic Energy, Government of India, for the Regional
Centre for Accelerator-based Particle Physics (RECAPP), Harish-Chandra
Research Institute. They also acknowledge the hospitality of the
Department of Theoretical Physics, Physical Research Laboratory (PRL), 
Ahmedabad, while the work was in progress. US would like to thank Prof. R. Cowsik, Director, McDonnell Center for the Space Sciences, Washington University in St. Louis, for arranging his visit as the Clark Way Harrison visiting professor, where this work has been completed.

\end{document}